\documentclass[%
twocolumn,
 amsmath,amssymb,
 aps, physrev,
]{revtex4-2}

\usepackage{graphicx}
\usepackage{dcolumn}
\usepackage{bm}
\usepackage{amsmath}

\usepackage{hyperref}
\usepackage{subcaption}
\usepackage[]{natbib}
\usepackage{ulem}

\newcommand{\DA}{D_{\rm A}}
\newcommand{\DM}{D_{\rm M}}
\newcommand{\DL}{D_{\rm L}}
\newcommand{\DHub}{D_{\rm H}}  
\newcommand{\DV}{D_{\rm V}}
\newcommand{\rd}{r_{\rm d}}

\begin{document}


\title{\textbf{Cosmology-Independent Constraints on the Etherington Relation and SNeIa Absolute Magnitude Evolution from DESI-DR2}}

\author{Sourav Das}
 \email{sourav.das@iucaa.in}
\author{Surhud More}%
 \email{surhud@iucaa.in}
\affiliation{%
The Inter-University Centre for Astronomy and Astrophysics, Post Bag 4, Ganeshkhind, Pune 411007, India
}%
\altaffiliation{%
Kavli Institute for the Physics and Mathematics of the Universe (WPI), 5-1-5 Kashiwanoha, Kashiwa-shi, Chiba 277-8583, Japan
}

\author{Shadab Alam}
\affiliation{
Tata Institute of Fundamental Research, Homi Bhabha Road, Mumbai 400005, India
}%

\date{\today}

\begin{abstract}
We carry out a test of the fundamental Etherington relation (cosmic distance duality relation) which relates the luminosity distance $\DL$ and angular diameter distance $\DA$ in metric theories of gravity. We use the latest measurements of the angular diameter distance as a function of redshift from the Dark Energy Spectroscopic Instrument Data Release 2 (DESI-DR2) and the luminosity distance from a variety of compilations of Supernovae of Type Ia (SNeIa). Our results indicate that these measurements are statistically consistent with the Etherington relation. In addition to providing a confirmation of the underlying assumptions of the Etherington relation, i.e., the metric nature of gravity, Lorentz invariance and photon number conservation, our results are also a stringent test of any residual systematic effects. We interpret the absence of evidence of any deviation from this relation to constrain the evolution of the absolute magnitude of SNeIa to $dM/dz = 0.07 \pm 0.07$ over and above the systematics that are already accounted for in the SNeIa analyses. We discuss how the Etherington relation can be used to constrain systematic parameters in the analyses of dynamical dark energy using geometric probes, to make it more robust against systematic effects.
\end{abstract}

\maketitle


\section{Introduction}

Distance measurements are fundamental to observational cosmology, especially to
connect various astrophysical observations to theoretical models of the
Universe. Distance redshift relations provide geometric constraints on the
cosmological models and have been crucial to understand the dynamics of the
expansion of the Universe.

In an expanding Universe, the observed flux of an astrophysical object is related
to its intrinsic luminosity through the luminosity distance such that
$F=L/(4\pi\DL^2)$. The dependence of $\DL$ on redshift account for the area of
the sphere over which the photons from an isotropic source get distributed, the
amount by which they get redshifted, and the amount by which their arrival times
between successive photons get stretched, as they travel through the Universe.
On the other hand, the angular diameter distance $\DA$ relates the observed
angular size of an object to its physical size, $S$, via $\theta=S/\DA$. The
dependence of $\DA$ on redshift is a result of the convergence of photons
arriving from the endpoints of an astrophysical object towards the observer.

The luminosity distance corresponds to the angle in which a bundle of light rays
spreads out from a source and the area it occupies at the observer. On the other
hand, the angular diameter distance corresponds to the area at the source and the
solid angle it subtends at the observer.  If photons travel along null
geodesics, local Lorentz invariance holds, and photon numbers are conserved,
these solid angles and areas are related to each other.

If photons travel along null geodesics, local Lorentz invariance holds, and
photon numbers are conserved, \cite{doi:10.1080/14786443309462220} present a
reciprocity relation which connects these solid angles and areas. The ratio of
the solid angle ($d\Omega_{\rm g}$) of a bundle of ray emanating from a source
to an area at the location of the observer ($dS_{\rm g}$) and the ratio of
the solid angle ($d\Omega_{\rm o}$) formed by a bundle of light rays converging
towards an observer to the extended area in the source ($dS_{\rm o}$) from which they
emanate \citep{Ellis2007}, such that
\begin{align}
\frac{d\Omega_{\rm o}}{dS_{\rm o}} = \frac{d\Omega_{\rm g}}{dS_{\rm g}}(1+z)^2 \,.
\end{align}
There is no direct way to measure $d\Omega_{\rm g}$, but it is related to the
flux, energy per unit area per unit time, and hence the luminosity distance. The
left-hand side, on the other hand, is related to the angular diameter distance.
The reciprocity relation thus translates to a relation between the angular
diameter distance and the luminosity distance such that
\begin{align}
    \DL = (1+z)^2 \DA\,.
    \label{eq:etherington_relation}
\end{align}
The above equation has been referred to in the literature via various names: the
cosmic distance duality relation, the Tolman test, the Etherington relation or
the reciprocity relation. Tests of the validity of this relation have also been
called the cosmic transparency test in the literature. We will refer to this
relation as the Etherington relation throughout this paper.

Over the last few years, there have been multiple attempts at testing the
Etherington relation observationally. Standardizable candles such as Supernovae
of Type Ia (SNeIa)\footnote{We will use SNIa to denote a single Type Ia
supernova.} have been used to measure luminosity distances to the host galaxies
of these SNeIa for cosmological purposes. The shape of the light curve of SNeIa
can be used to standardize the luminosity of the SNeIa, and convert the flux
measurements into a luminosity distance \citep{Phillips, SALT2, Jha2007}. The
luminosity distance measurements from SNeIa provided some of the early evidence
for the presence of an accelerated expansion of the Universe
\citep{Perlmutter1999, Schmidt1998, Riess_1998}. When combined with measurements
of the cosmic microwave background \citep[see e.g.,][]{Spergel2003, Dunkley2008,
PlanckCollaboration2020}, and measurements of matter density parameter in
clusters \citep[see e.g.,][]{Allen2002}, these measurements established the
standard cosmological model \citep[see e.g., ][]{Knop2003, Kowalski2008}.

The standard cosmological model states that 95 percent of the total energy
density of the Universe today consists of two mysterious components, dark matter
and dark energy, with the rest accounted by ordinary matter. Dark energy is
often modeled as a cosmological constant with an equation of state parameter $w=-1$
which connects the pressure and energy density, such that $p=w\rho c^2$. The
search for any dynamical nature for dark energy is ongoing.

Given the importance of the luminosity distances from SNeIa to the cosmological
model, efforts have been made for testing the reliability of these estimates
using the Etherington relation. Using a variety of estimates of the angular
diameter distance from different astrophysical sources such as FRIIb radio
galaxies, compact radio sources, Gamma Ray bursts and X-ray clusters at
cosmological distances, Einstein radii of strong lenses, or $H(z)$ measurements
from luminous red galaxies, the Etherington relation has been used as a test for
systematics in the luminosity distances from SNeIa \citep{bassettkunz2004a,
bassettkunz2004b, uzan2004, jackson2008, avgoustidis2009, avgoustidis2010,
holanda2011, shuo_nan2011, Lima_Cunha2011, fu_wu_yu_2011, li_wu_yu2011,
holanda2012, liang_li_wu_2013, holanda2016, fu_lu_2017, ruan_melia_zhang_2018,
Renzi2022, yang_fu_xu_2024}.

The various astrophysical sources can result in a number of systematics related
to the determination of the angular diameter distances. In \citet{more2009}, it
was therefore suggested that the angular diameter distances from the baryon
acoustic feature (BAF, also referred to as Baryon acoustic oscillations or BAO)
be used in order to carry out the Etherington test \citep[see also][]{Nair2012,
moreniikura2016, kai_avgoustidis_2015, xu_wang_zhang_2022}. The BAF arises at a
distance corresponding to the sound horizon, $r_{\rm d}$, at the redshift drag
epoch when the baryons stop being dragged by the photons after decoupling
\citep[see e.g.,][]{EisensteinHu1998}. This length scale is preserved through
the Universe's expansion in the correlation function of the matter distribution
and can be measured by using biased tracers of the matter distribution such as
galaxies and quasars.

Observationally, the BAF is imprinted as a bump in the galaxy and quasar
two-point cross-correlation statistics at a characteristic angular scale or as
oscillations of a characteristic angular frequency in their power
spectra \citep{Eisenstein_2005, boss, eboss, desidr1}. The ratio of the angular
scale at which the feature occurs to the precise measurements of the angular
scale of the sound horizon imprinted in the cosmic microwave background,
provides a convenient measure of the ratio of the two angular diameter
distances.

BAF measurements are typically reported as constraints on cosmological distances
scaled by $\rd$. Transverse measurements constrain $\DM(z) / \rd$, where $\DM$
is the transverse comoving distance, radial measurements constrain $\DHub (z) /
\rd$, where $\DHub = c/H(z)$, and isotropic analysis estimates the volume-averaged
distance $\DV(z) / \rd$ where $\DV(z) = (z\DM^2\DHub)^{1/3}$.

Since these quantities are inferred relative to $\rd$, any conversion to $\DA$
requires either adopting a value (or a prior) for $\rd$ or treating $\rd$ as a
nuisance parameter to be marginalized over. In our analysis, we convert the DESI
BAO observables into $\DA(z)$ at the reported redshifts using the transverse
comoving distance measurements where available, and we eliminate any need for
$\rd$ to be estimated, by taking the ratio of the angular diameter distances at
different redshifts.

The latest results from the BAF studies come from the Dark Energy Spectroscopic
Instrument survey (DESI), where the BAF measurements have been presented for 9
different samples at different redshifts using Data Release 2 (DESI-DR2). The
unprecedented accuracy of these results, the large redshift range they cover,
allow a stringent test of the standard cosmological model, in particular the
variation of the equation of state parameter for dark energy
\citep{Chevallier2001, Linder2003}. When combined with the CMB measurements and
SNeIa data (marginalized over absolute magnitude uncertainties), the DESI-DR2
measurements claim a tentative but growing evidence for a detection of dynamical
dark energy \citep{DESI_DR2, Lodha_2025}.

Given the importance of this tentative evidence and its importance for the
standard cosmological model, these results have received a large scrutiny from
the scientific community. While the existence of dynamical dark energy has
spawned a number of theoretical models \citep[see e.g.,][]{Silva_2025,
PanSupriya_2025, Mishra_2025}, the consistency of data combinations of
BAF, CMB and SNeIa has also received scrutiny, as neither dataset on its own
shows evidence for dynamical dark energy \citep[see e.g.,][]{popovic_shah_2025,
Ormondroyd_2026}. To add to the intrigue, the degeneracy contours from the
DESI-DR2  point towards the $w_0=-1$, $w_a=0$, the standard cosmological model
prediction.

Recently, \citet[][AM25 hereafter]{afroz2025hintinconsistencybaosupernovae} have
used the Etherington relation and claimed an inconsistency between data sets
when the BAF from DESI-DR2 results are analyzed in conjunction with the
Pantheon+ data \citep{Scolnic_2022}. If true, this would indicate a breakdown of
the assumptions underlying the Etherington relation, or it could also be a
possible evidence for systematics in the data, calling in to question the
validity of the evidence for dynamical dark energy. When allowing for a
violation of the Etherington relation, the authors claim the hint for dynamical
dark energy disappears. In this paper, we critically examine the claim for the
violation of the Etherington relation using the various SNe data sets used in
the DESI-DR2 cosmological analysis \citep{
Scolnic_2022,Vincenzi_2024,Rubin2023}.

Our analysis follows the methodology and techniques established in
\cite{more2009, moreniikura2016} and carries out robust tests of the Etherington
relation over a wide redshift range $0.41<z<1.58$ and thus tests for any
inconsistency due to systematics in either the BAF or the SNeIa measurements,
which could hamper a successful combination of these measurements for obtaining
constraints on the dark energy equation of state parameter.

This paper is organized as follows. We discuss the data being used in our
analysis to compute the distances in section \ref{sec:data}, and the methodology
implemented by us in section \ref{sec:method}. Finally, we present the results
in section \ref{sec:results} and discuss the results in section \ref{sec:discussion}.

\section{Data}\label{sec:data}

The Etherington test require measurements of the angular diameter distance from
standard rulers and the luminosity distance from standard candles at a given
redshift. Obtaining both measurements from individual objects is extremely
challenging. Therefore we resort to ensemble measurements of these distances
carried out in the recent past.

\subsection{Type Ia Supernovae}

SNeIa are thermonuclear explosions of white dwarfs and are
well-established standardizable candles \citep{Phillips, 1993ApJ...405L...5B}.
The decline of the B-band light curve 15 days after maximum, $\Delta m_{15}(B)$,
or equivalently the light-curve ``stretch,'' correlates with its intrinsic peak
luminosity. After correcting each SNIa for stretch, the observed flux at the peak
can then be used to obtain luminosity distance to the host galaxy of the SNIa.
The supernovae fluxes can also be affected by the dust within the host galaxy,
which needs to be corrected for before the luminosity distances can be used for
cosmological analyses.

Although SNeIa are standardizable, an absolute calibration is still required
to set their luminosity zero-point. Traditionally, this calibration is provided
by the cosmic distance ladder, combining geometric anchors and Cepheid
variables \citep{Riess_2022}. The Etherington test we propose bypasses this
absolute calibration by considering ratios of luminosity distances at two
redshifts, which depend only on relative distances. In our analysis, we use
various SNeIa compilations given that we wish to carry out the Etherington test
over a wide range of redshifts.

The Joint Lightcurve Analysis \citep[JLA;][]{Betoule_2014} is a compilation of
740 SNe primarily obtained from the Sloan Digital Sky Survey II
\citep[SDSS-II;][]{SDSSII} and the Supernova Legacy Survey
\citep[SNLS;][]{SNLS}. A smaller fraction comes from the low redshift surveys
as well as high redshift discoveries from the Hubble Space Telescope. The
photometry of SNe discovered from these surveys was homogenized and fit using
the SALT2 light curve model \citep{SALT2}. The distance modulus to the host
galaxy of every SNIa is given by,
\begin{align}
\mu = 5 \log_{10}\left( \frac{d_L}{10 {\rm pc}} \right) = m_B^* - (M_B -\alpha X_1+ \beta C) \,,
\end{align}
where $d_L$ is the luminosity distance, $m_B^*$ is the peak apparent B-band
magnitude of the SNIa, $\alpha$ corresponds to standardization correction due
to the stretch parameter $X_1$ that describes the SNIa light curve, while $\beta$
corresponds to the correction for dust extinction applied depending upon the
color $C$ of the SNIa. The symbol $M_B$ corresponds to the absolute magnitude
of the SNIa which can have additional dependencies on the type of the host
galaxy.

The parameters $m_B^*$, $X_1$, and $C$ are determined from the fit to the
time-spectral data obtained from the SNeIa. The joint light curve fitting
results in the measurements of the above parameters in a correlated manner, and
the covariance matrix of the resulting measurements is also provided in the
catalogs.

The spectroscopically confirmed SNeIa discovered from Pan-STARRS1 Medium Deep
Survey, have been combined with catalogs of previously identified supernovae
from SDSS-II and SNLS to form the Pantheon sample \citep{Scolnic_2018}
consisting of $1048$ SNeIa. The data from this sample includes bias corrections
due to photometric selection effects and light curve fitting to noisy data
\cite{Kessler2017}. The Pantheon+ compilation which we use in this paper
incorporates re-analyzed and newly calibrated photometry for a total of $1701$
spectroscopically confirmed SNeIa across the redshift range $0.001<z<2.26$
\citep{Brout_2022}. This dataset significantly improves upon its predecessor
by applying a consistent photometric calibration across different surveys,
enhancing the statistical precision and minimizing systematic uncertainties in
cosmological parameter estimation.

The Union3 compilation is a larger sample of SNeIa that builds upon its
previous versions and comprises of 2087 SNeIa with standardized light-curve
fitting and homogenized calibration \citep{Rubin2023}. We also use this dataset
due to its broad redshift coverage and inclusion of many previously published
SNeIa.

Finally, in all these compilations, systematic uncertainties can arise while
homogenizing the SNeIa detected from different surveys and calibrating the
photometric systems used to obtain their light curves. Therefore, we also use
the $\sim1600$ SNeIa which were classified photometrically using data from the
Dark energy survey \citep{DESY5}. A dedicated follow up campaign was used to get
the redshifts of the host galaxies. Compared to compilations such as Pantheon+,
given the depth of the Dark energy survey, this compilation has about 5 times
larger sample of SNeIa at redshifts $z>0.5$.

\subsection{BAO data}

We utilize the distances inferred from the analysis of the BAF measurements
from spectroscopic data obtained by the DESI experiment in its Data Release 2
(DR2) \citep{DESI_DR2}. The DESI DR2 enables measurements of the
autocorrelation of various samples of galaxies and of quasars as large scale
structure tracers at a wide range of redshifts. These clustering measurements
show significant evidence for the presence of the baryon acoustic feature,
which can be used as a standard ruler to infer distance measurements to the
effective redshifts of the various samples.

Although the intrinsic clustering of the tracers is expected to be isotropic,
the peculiar velocities of these tracers result in an anisotropy in the
measurements of the clustering signal. The anisotropy arises from two effects.
Firstly, the positions and redshifts of galaxies are converted to distances in
three dimensions in the context of a fiducial standard cosmological model.
Peculiar velocities, on small scales due to virial motions of galaxies within
dark matter halos, i.e. the finger-of-god effect \citep{jackson1972, huchra1988}, and on large scales due to
motion of galaxies towards large scale overdensities, i.e., the Kaiser effect \citep{kaiser1987, hamilton1992},
result in anisotropic distortions in the separations of galaxies.
The difference between the fiducial cosmological model and the true underlying
model for the Universe further result in anisotropic distortions, i.e. the
Alcock Paczynski effect \citep{alcock_paczynski_1979}.

Given these distortions, for most samples, analysis of the DESI DR2 data includes measurements of the clustering of the corresponding tracers in a two-dimensional plane of separations along the line-of-sight and in the plane of the sky. The BAF in the line-of-sight provides measurements of the $D_{\rm H}/r_{\rm d}$, while that in the plane of the sky provides measurements of the transverse comoving distance $D_{\rm M}/r_{\rm d}$. Where the signal-to-noise ratio is not enough to detect the BAF separately in the two directions, the clustering is combined in the two directions to infer a combination of the two measures, $D_{\rm V}/r_{\rm d}=[D_{\rm H}D_{\rm M}^2 z]^{1/3}/r_{\rm d}$. The particular combination is a dimension-weighted average of the two distance measures.

Given that the Etherington test requires inferred values of $D_{A}=D_{\rm M}/(1+z)$, we only use those samples where the signal-to-noise ratio was high enough for the BAF to be detected. In particular, we will use only the inferences of distances from those tracers that provide $D_{\rm M}$ as listed in Table~\ref{tab:table_bao_data}. The BAF inferred distance posteriors marginalized over $D_{\rm H}$ are tabulated as mean and a 1-$\sigma$ error.

\section{Method}\label{sec:method}

Our aim is to test the validity of the Etherington relation as a function of redshift. In order to carry out a model independent test which will be insensitive to the uncertainties in the absolute magnitudes of SNeIa or to the comoving sound horizon at the drag epoch, we compute the ratio of $\DL/\DA$ at a given redshift to the ratio at a reference redshift. We define the quantity ${\cal R}(z, z_{\rm ref})$ such that
\begin{align}
    {\cal R}(z, z_{\rm ref}) \equiv \frac{\DL(z)\DA(z_{\rm ref})}{\DA(z)\DL(z_{\rm ref})}
\end{align}
The Etherington test now becomes a consistency test between ${\cal R}(z, z_{\rm ref})$ and the ratio $\left[(1+z)/(1+z_{\rm ref})\right]^2$.

In order to compute ${\cal R}$, we require measurements of the luminosity distance $\DL$ and the angular diameter distance $\DA$ at similar redshifts. Various approaches have been used in the literature for this purpose. In some of the earlier works, the luminosity distance measurement from the closest supernova to the redshift at which the BAF distance was quoted has been used to carry out the Etherington relation. This approach introduces quite a bit of uncertainty in the Etherington test arising from the statistical errors on the inference of $\DL$ from a single supernova.

Another approach is to use a certain redshift interval $\Delta z$ within the BAF redshift in order to select supernovae to determine the value of the luminosity distance. Such an approach brings in its own dependence on the width of the redshift interval. A large width of the redshift interval will lead to averaging the luminosity distances over a range of redshifts where it varies significantly. On the other hand a smaller redshift interval will lead to larger statistical errors.

Yet another approach is to carry out an interpolation in order to infer the luminosity distance at the relevant redshift where the BAF has been used to obtain the angular diameter distance. Various interpolation techniques which either use Gaussian processes or neural networks have been used in order to infer the luminosity distance at the relevant distance in the literature. In our analysis, we follow the strategy from \cite{more2009, moreniikura2016} in order to interpolate the value of the luminosity distance as it follows a parallel strategy that is adopted in the BAF analyses.

In the BAF analyses, the tracers often span a range of redshifts. In this case, a fiducial cosmological model is adopted in order to measure the clustering signal. The difference between the fiducial cosmological model and the underlying true model would lead to a shift in the location of the BAF. This shift is then interpreted as a ratio of the inferred angular diameter distance to the angular diameter distance in the fiducial cosmological model.

We adopt a fiducial flat $\Lambda$CDM cosmological model for computing the corresponding distance modulus as a function of redshift with parameters $\Omega_{m,0}=0.3$ and $H_0=100 {\rm km\,s^{-1}\,Mpc^{-1}}$. In order to fit the SNeIa data (either the distance modulus $\mu(z)$ or the SNeIa peak magnitudes), we use a parametric model which consists of a quadratic polynomial on top of the fiducial distance modulus-redshift relation\footnote{We have checked that using a cubic polynomial instead does not change our conclusions about the consistency of the Etherington relation.}. Thus our total model for the distance modulus as a function of redshift is given by
\begin{align}
    \mu(z) = \mu_{\rm fid}(z) + A(z-z_{\rm ref})^2 + B(z-z_{\rm ref})
\end{align}
This approach allows us freedom in the distance modulus to incorporate any deviations from the fiducial cosmological model we adopt. Note that there is no constant deviation that we have included in the above relation. Any such constant would drop out when we subtract the distance modulus between two redshifts which are required to compute the ratio of the luminosity distances.

We implement our likelihood in the cosmology inference pipeline {\sc COBAYA} and sample the posterior distribution for the parameters $A$ and $B$ by using a Markov Chain Monte Carlo \cite{Torrado_2021}. The posterior distribution of these parameters can be used to obtain a model for $\mu(z)$ for any of our SNeIa data compilations, we can obtain the difference between the distance moduli at any two redshifts. Our fit will also include any correlated errors in $\Delta\mu$ across different redshift bins.

Given the different redshift ranges spanned by the data used for the BAF
measurements, we also restrict ourselves to the same redshift ranges for the SNe
data. The minimum redshift we consider is $0.41$, corresponding to the redshift
bin for the BAF measurements from the LRG1 tracer. The maximum redshift depends
upon the SNe compilations, since none of them extend out to the maximum
redshifts for the quasars used for the BAF measurement. For every compilation we
list the number of SNe used, their selection based on the redshift range, and
whether the compilation provides the distance moduli or the supernova apparent
magnitudes, is listed in Table~\ref{tab:table_sn_data_bins}.

\begin{figure*}
    \centering
    \begin{subfigure}{0.48\textwidth}
        \centering
        \includegraphics[trim={0 0 0 1cm},clip,width=\linewidth]{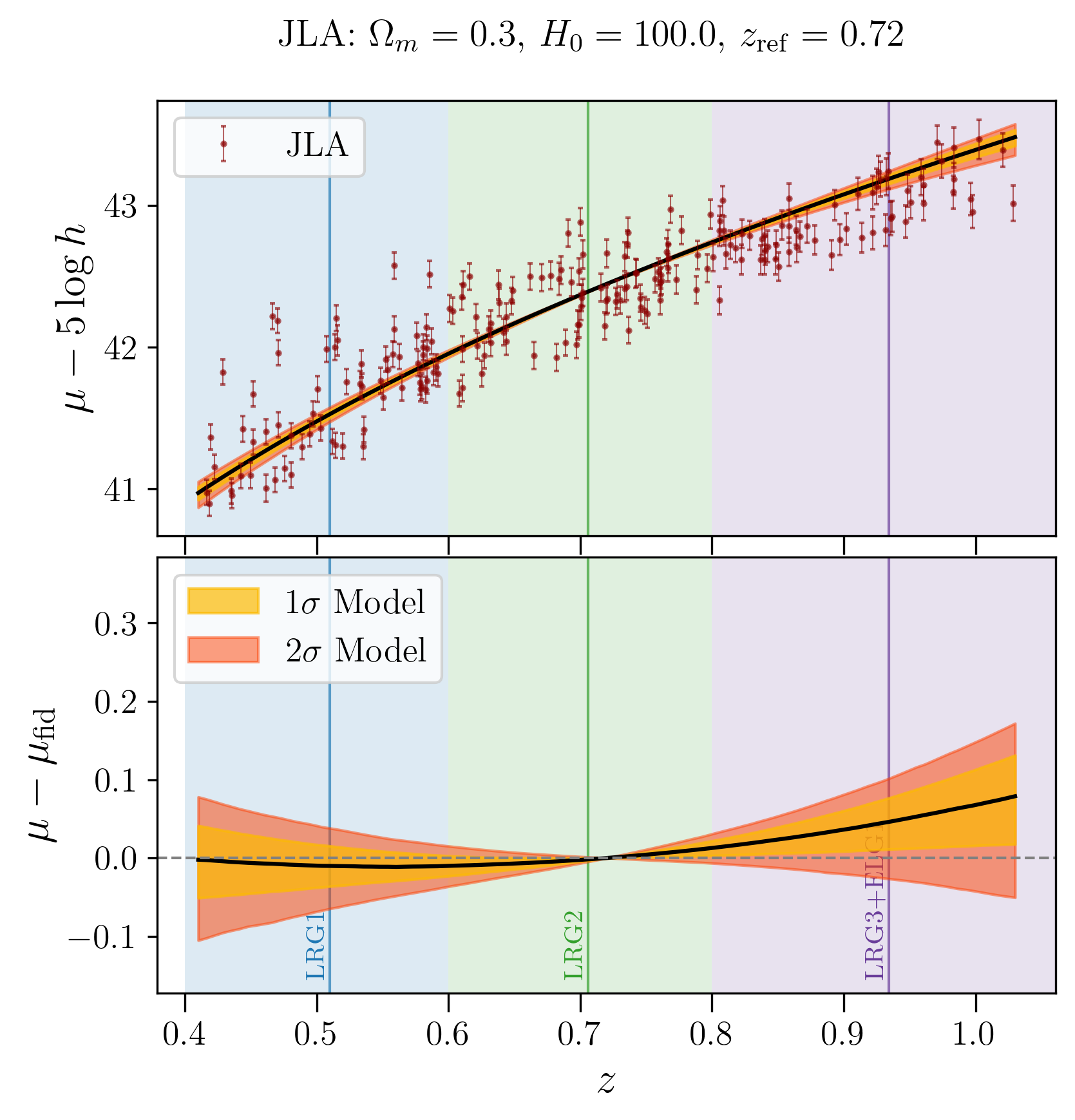}
        \caption{JLA}
        \label{fig:2panel_jla}
    \end{subfigure}
    \hfill
    \begin{subfigure}{0.49\textwidth}
        \centering
        \includegraphics[trim={0 0 0 1cm},clip,width=\linewidth]{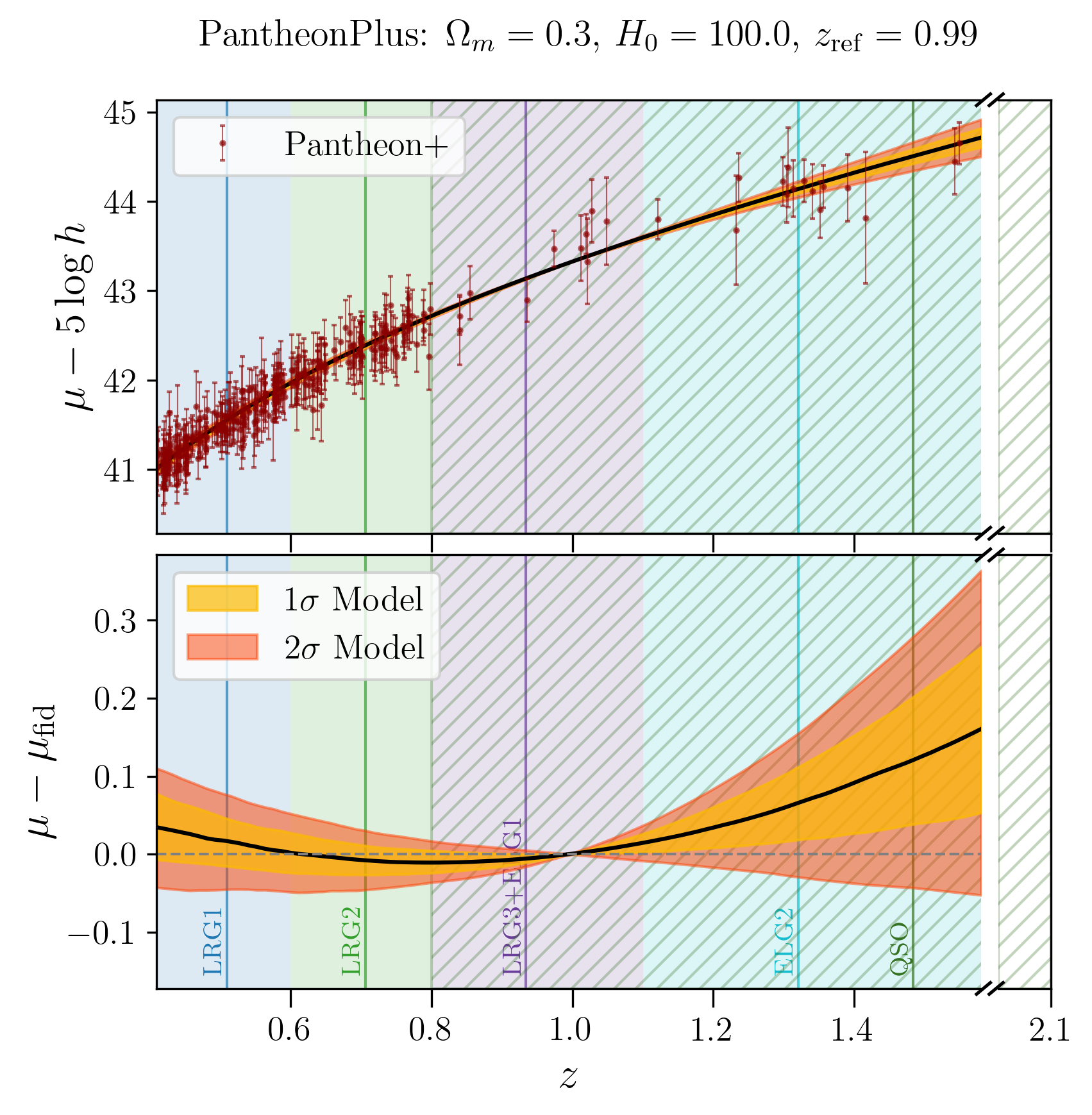}
        \caption{PantheonPlus}
        \label{fig:2panel_pantheonplus}
    \end{subfigure}


    \begin{subfigure}{0.49\textwidth}
        \centering
        \includegraphics[trim={0 0 0 1cm},clip,width=\linewidth]{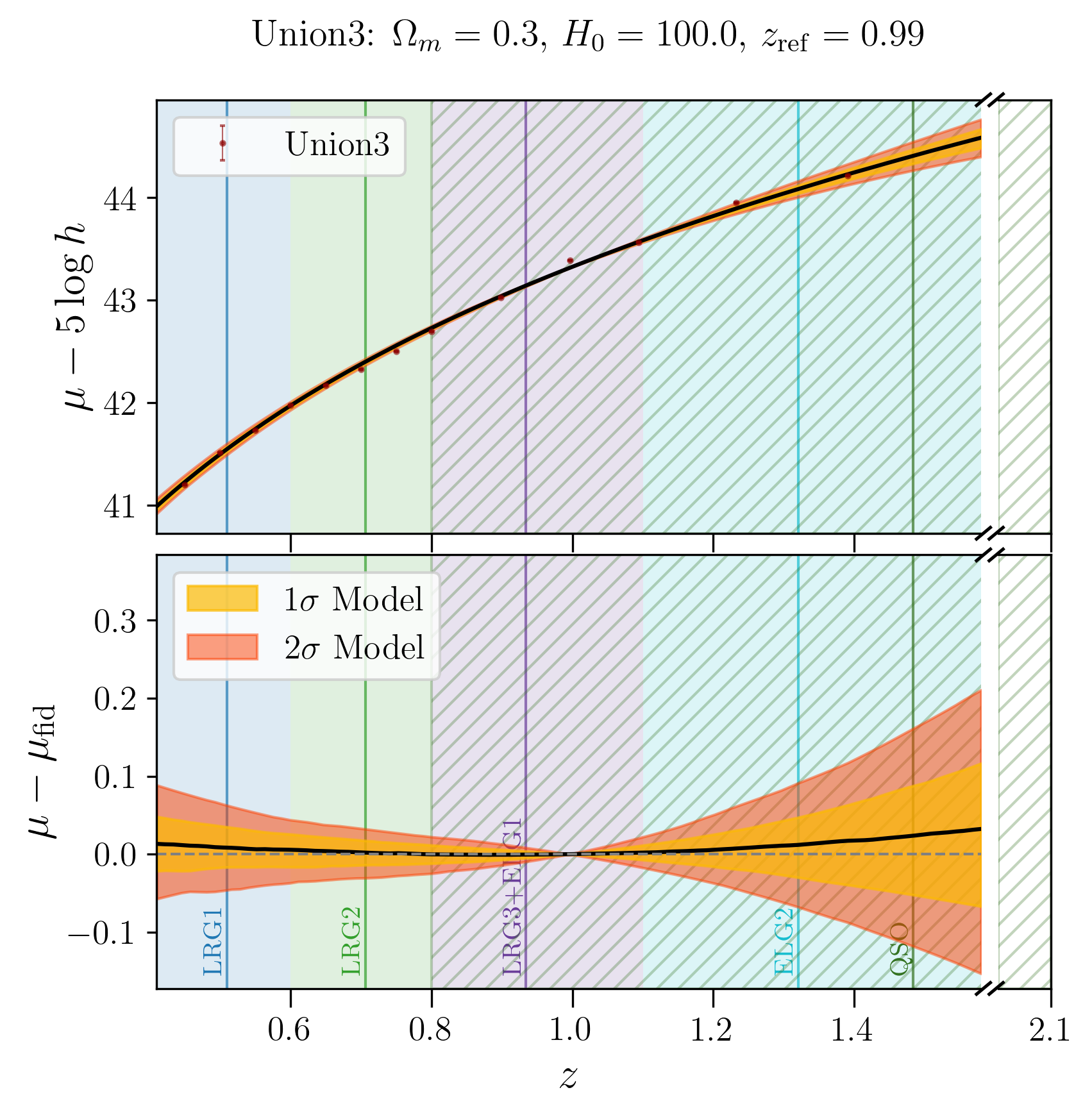}
        \caption{Union3}
        \label{fig:2panel_union3}
    \end{subfigure}
    \hfill
    \begin{subfigure}{0.48\textwidth}
        \centering
	    \includegraphics[trim={0 0 0 1cm},clip,width=\columnwidth]{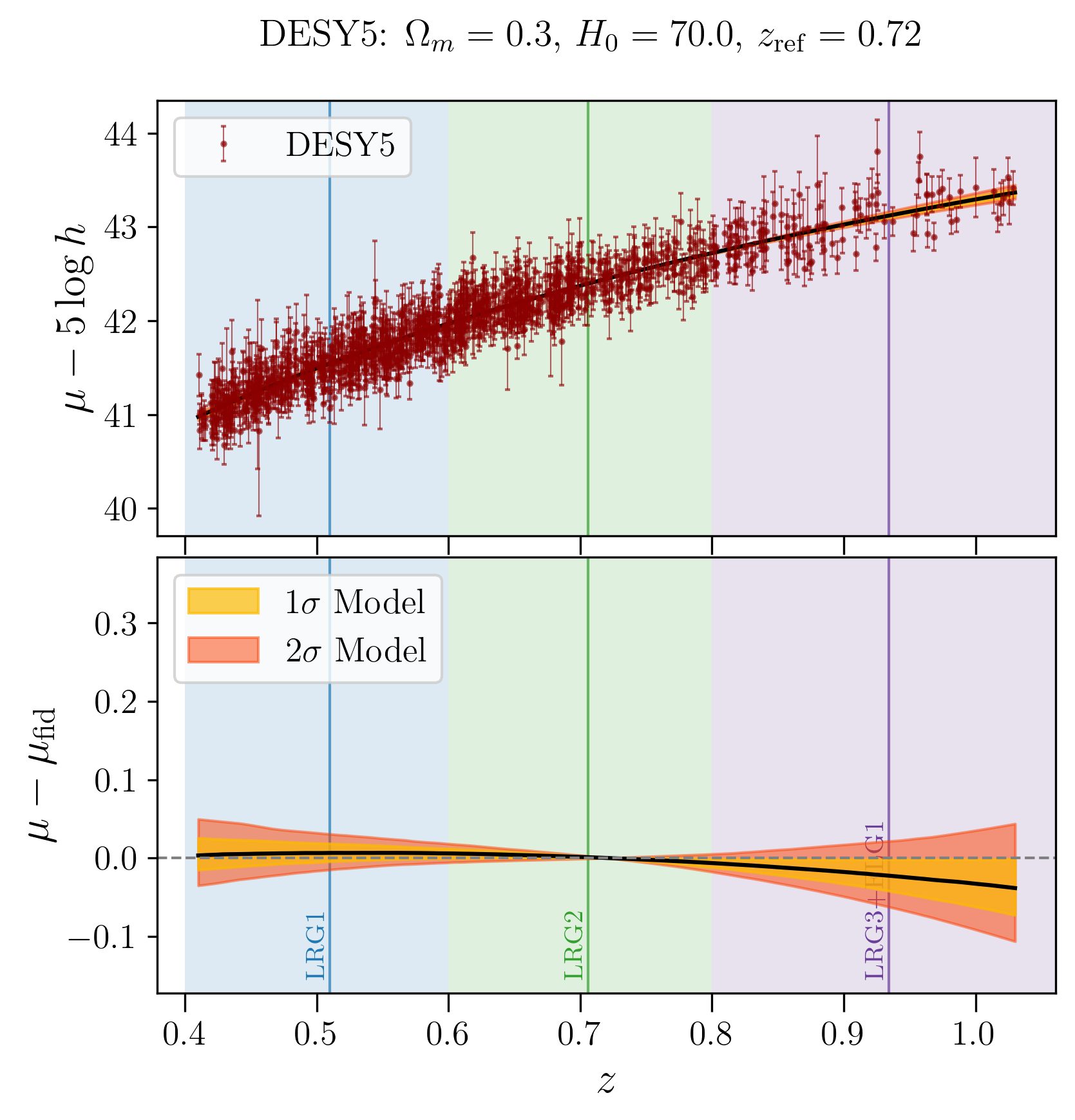}
        \caption{DESY5}
	    \label{fig:2panel_desy5}
    \end{subfigure}
    \caption{
    The data points with errorbars in different panels show the distance modulus
    corresponding to SNeIa compiled by (a) JLA, (b) Pantheon+, (c) Union3, and
    (d) DESY5 as indicated at the top of each panel. In each panel, {\it top
    subpanel} shows the distance modulus $\mu - 5\log{h}$ obtained from SNeIa,
    {\it bottom subpanel} shows the corresponding residuals which are fit with a
    quadratic function. The 68 and 95 credible regions for the residuals are
    shown with the shaded region. 
    }
    \label{fig:combined_plots}
\end{figure*}

\begin{table}
\centering
\caption{The redshift binning scheme for various SNeIa compilations}
\label{tab:table_sn_data_bins}
\begin{tabular}{|c|c|c|c|c|}
\hline
\textbf{Compilation}      & $z_{\rm min}$ & $z_{\rm max}$ & $N_{\rm SN}$ & Data \\ \hline
\textbf{JLA}          & 0.41       & 1.03      & 205 
& $m_{\rm B}$             \\
\textbf{Pantheon}     & 0.41       & 1.58          & 268 
& $m_{\rm B}$             \\
\textbf{PantheonPlus} & 0.41       & 1.58        & 291         & $m_{\rm B}$           \\
\textbf{Union3}       & 0.41          & 1.58      & 13           & $m_{\rm B}$             \\
\textbf{DESY5}        & 0.41       & 1.03       & 1078         & $\mu$             \\ \hline
\end{tabular}
\end{table}

\begin{table}
    \centering
    \caption{DESI DR2 BAF inferences of $\DM/\rd$ with errors \citep{DESI_DR2}}
    \label{tab:table_bao_data}
    \begin{tabular}{|c|c|c|l|l|}
        \cline{1-5}
        \textbf{$z$} & \textbf{$\DM/\rd$} & $\sigma_{\DM/\rd}$  & \textbf{Tracer} & \textbf{redshift}  \\ \cline{1-5}
        0.510 & 13.59 & 0.17 & LRG1      & $0.4{-}0.6$  \\
        0.706 & 17.35 & 0.18 & LRG2      & $0.6{-}0.8$  \\
        0.934 & 21.58 & 0.16 & LRG3+ELG1 & $0.8{-}1.1$  \\
        1.321 & 27.60 & 0.32 & ELG2      & $1.1{-}1.6$  \\
        1.484 & 30.51 & 0.76 & QSO       & $0.8{-}2.1$  \\ \cline{1-5}
    \end{tabular}

\end{table}

\section{Results}\label{sec:results}

We begin by presenting our results for the distance modulus as a function of redshift with the 5 different supernova datasets we use. The top panel of Figs.~\ref{fig:2panel_jla}--\ref{fig:2panel_desy5}, show the distance modulus, while the bottom panel show the residual $\mu - \mu_{\rm fid}$, as a function of redshift. The shaded regions in each of these panels indicate the redshift ranges over which BAF measurements have been performed given the redshift range of the included supernovae in each of these compilations. For samples which span redshift range greater than 1.5, we also include the BAF measurements from quasars from DESI. This redshift range is indicated as a shaded region. The corresponding effective redshifts where the BAF measurements are used to infer the angular diameter distance are shown with vertical solid lines.

In each of the panels of these figures, the shaded region indicates the 68 and 95 percent credible interval for the corresponding quantity. We note that the residuals in the inferred distance modulus and the fiducial distance modulus can be fit by a simple quadratic form as we had assumed. At the reference redshift the residual goes to zero by construction, and we have verified that the error bar and the covariance in the measurements of $\Delta \mu$ between two different redshifts are not affected by the exact choice of the reference redshift.

In Fig.~\ref{fig:2panel_jla}, we note that the data seem more dispersed than the underlying model, as the intrinsic scatter is determined as a separate parameter of the covariance matrix. Amongst the different compilations DESY5 shows the best residuals, given the large number of SNeIa. However, note that a large number of these are photometrically identified supernovae. As the residuals are also computed with respect to the same underlying fiducial model, they also show the differences between the measurements, and methodologies used to process the supernovae. Each of these compilations can thus independently inform us about possible systematics.

\begin{figure}
    \centering
    \includegraphics[trim={0 0 0 7.5mm},clip,width=\columnwidth]{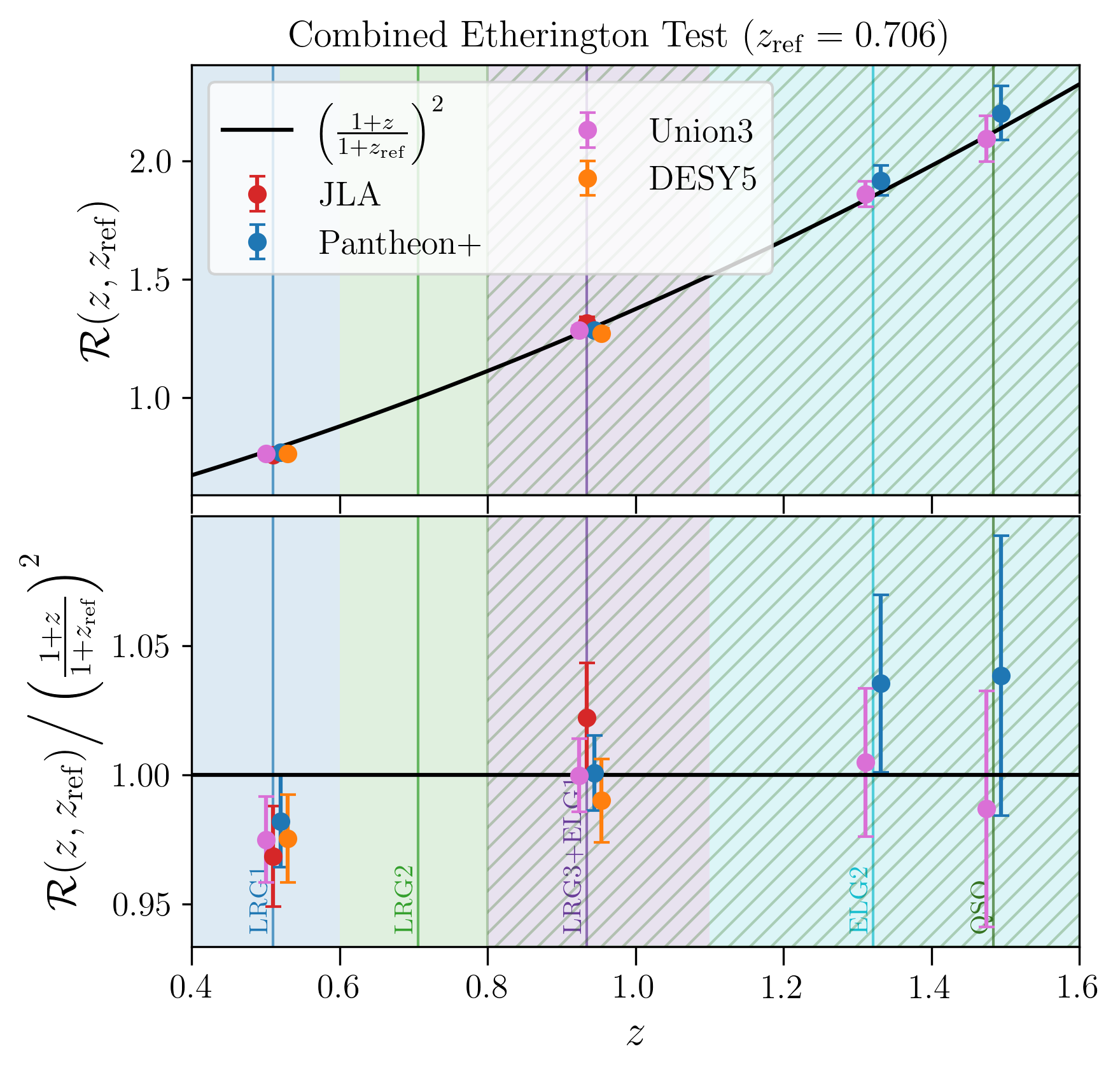}
    \caption{The validity of the Etherington relation as seen using the quantity ${\cal R}(z, z_{\rm ref})$, with $z_{\rm ref}=0.706$ using $\DL$ measurements from a variety of SNeIa compilations and $\DA$ measurements from DESI DR 2.}
    \label{fig:eth_combined_plot}
\end{figure}

\begin{figure}
    \centering
    \includegraphics[trim={0 0 0 1cm},clip,width=\columnwidth]{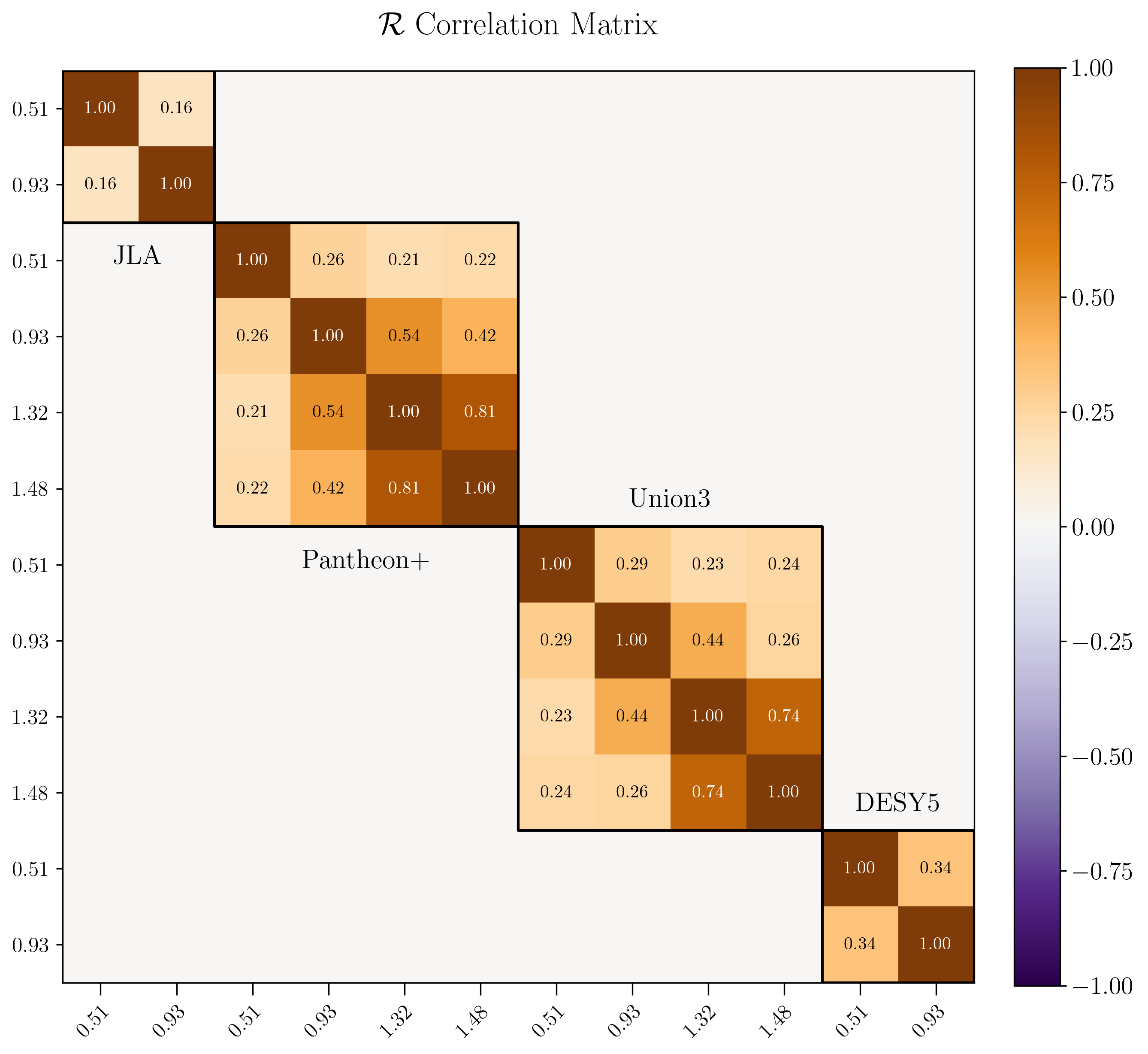}
    \caption{Cross-correlation matrix for ${\cal R}(z, z_{\rm ref})$ for various SNeIa samples. Note that the different samples should not be combined as they may contain data from same SNeIa analyzed under differing assumptions. We caution the reader that the block diagonal form of the cross-correlation matrix that appears here is for the economy of representation of all cross-correlation matrices in a single figure. Our analysis treats each one separately.}
    \label{fig:cross_corr}
\end{figure}

Given these measurements, we now perform the Etherington test, by determining the value of ${\cal R}$ at different redshifts where the BAF has been measured. For this purpose we chose $z_{\rm ref}=0.706$, corresponding to the LRG2 sample where BAF has been measured by the DESI collaboration. The data points and the errors in upper panel of Fig.~\ref{fig:eth_combined_plot}, show our inferred ${\cal R}(z, z_{\rm ref})$. These data can be compared with the expectation $(1+z)^2/(1+z_{\rm ref})^2$, which is shown as the solid black line. We observe that the data points obtained from our analysis closely tracks the expectation. We plot the ratio of ${\cal R}$ with this expectation in the bottom panel to clearly demonstrate any deviations from the expectation. Note that the errors on the data points are correlated with each other. We show the cross-correlation matrix for each of the points separately in Fig.~\ref{fig:cross_corr}.

We use these covariances to compute the consistency between the inferred values and the expectation. We find that all the different Supernovae data set that we test against the BAF results obey the Etherington test at various levels of consistency. Finally, we also fit a redshift dependent model to actively search for any residual redshift dependence of the form $(1+z)^{\alpha}$, and find $\alpha=0.09\pm0.07$ and $\alpha=0.07\pm 0.06$ when we fit the deviations of ${\cal R}$ using the Pantheon+ and Union3 samples, respectively.

Our results seem to be in apparent contradiction with the inconsistency pointed out between the supernovae and BAF measurements pointed out recently by AM25. There are a couple of notable differences between our methodologies. In this paper, we have used ratios between the luminosity distances and the angular diameter distances, instead of absolute values of these measurements. Any difference in the determination of the supernova calibration can cause offsets which are unrelated to the violation of the Etherington relation. Similarly any uncertainties in the absolute BAF calibration due to the sound horizon at redshift drag epoch, could also cause offsets unrelated to the validity of the Etherington relation. The kind of violation shown in fig. 4 of AM25 should have shown up even in the ratio ${\cal R}$, and we do not see any evidence for this in our analysis. As commented in the DESI DR2 analysis \cite{DESI_DR2}, while combining the SNeIa and the BAF dataset in the DESI analyses, the absolute calibration of the supernovae is entirely marginalized over and thus the SNe fit is performed in $h-$free variables, exactly in the same manner that we have performed our fits. In addition, AM25 use a single value of $r_{\rm d}$ instead of marginalizing over this parameter. This effectively neglects the covariance for the BAF inference, as the fit carried out within individual redshift bins can exaggerate any statistical fluctuations. Our analysis entirely avoids these issues altogether with the help of ratios, focussing on the violations of the Etherington relation explicitly.

In another recent work, \cite{Teixeira2025}, fit for the deviations of the Etherington relation using similar data set, and find evidence for $\DL/\DA(z) = 0.925(1+z^2)$. Such a deviation in the Etherington relation is expected due to constant systematic offsets, and the test we carry out with ${\cal R}$ cannot rule it out. However, according to \cite{Teixeira2025}, their preferred model is $\DL/\DA=(1+z)^{1.866}$. This will correspond to a 5 percent deviation in ${\cal R}$ from the expectation between the reference redshift $0.706$, and $z=1.5$. We see some marginal evidence for this in the bottom panel of Fig.~\ref{fig:eth_combined_plot} when using the Pantheon+ compilation of SNe results, but do not see such evidence in the Union3 compilation. This raises the possibility that within the systematic uncertainties related to different treatments of the SNe data, we do not see significant evidence for a breakdown in the Etherington relation.

\begin{figure}
    \centering
    \includegraphics[trim={0 0 0 8mm},clip,width=\columnwidth]{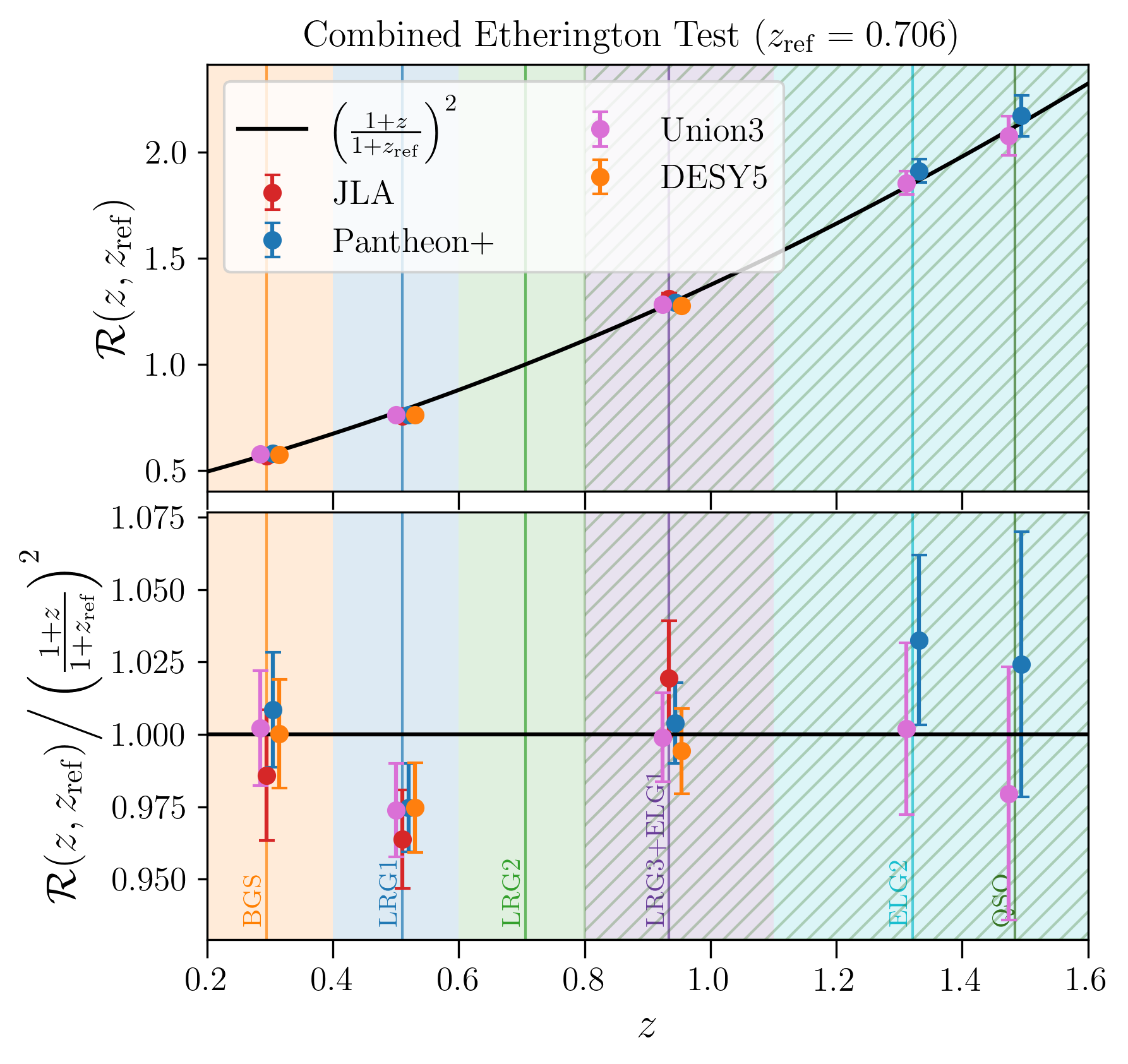}
    \caption{Same as Figure \ref{fig:eth_combined_plot}, but extended to lower redshifts to include the DESI Bright Galaxy Survey (BGS) measurement at $z=0.295$. Since the BGS data only provides the volume-averaged distance $D_V/r_s$, we extract the angular diameter distance by adopting a fiducial Planck 2018 $\Lambda$CDM cosmology for $D_H/r_s$. This introduces a mild model dependence to the lowest redshift data point, breaking the strict model independence maintained in the rest of the analysis.}
    \label{fig:eth_combined_plot_extra}
\end{figure}

Until now, we have tested the Etherington relation in a manner completely independent of the cosmological model. However, AM25 use one more data point from the BAF at low redshift from the use of the DESI Bright galaxy sample (DESI-BGS). The smallest uncertainty in the Etherington test carried out in AM25 corresponds to this particular redshift bin. However, the analysis of DESI-BGS sample allows the inference of the quantity $\DV$ instead of $\DA$, which is why it is not listed in Table~\ref{tab:table_bao_data}. As it is a combination of $\DHub$ and $\DA$, we need an independent estimate of $\DHub$. In what follows we use the best estimate of $\DHub$ from the cosmological model constrained with CMB data from Planck \citep{Planck_2018} in order to convert $\DV$ to $\DA$ following AM25. We carry out our entire analysis for ${\cal R}(z)$ including this extra data point. The corresponding results and fits are shown in Fig.~\ref{fig:eth_combined_plot_extra} and confirm that we still observe consistency with the Etherington relation. The corresponding p-values to exceed $\chi^2$ are $0.12, 0.29, 0.42, 0.36$ for the JLA, Pantheon+, Union3 and DESY5 samples, respectively. We get similar order of magnitude numbers for the p-values even if we use $\DHub$ based on the best DESI-DR2 flat $w_0-w_a$CDM cosmology. Given that we have not marginalized over any uncertainty in the determination of $\DHub$ while carrying out this test, implies that these are further conservative estimates of the p-values. Nevertheless, we do point out that we do see a very tentative difference (less than $2\,\sigma$) with the Etherington relation expectation at the redshift corresponding to LRG1 sample, and that this should be carefully attended to as the data allow further precise tests in the near future.

It may well be that there are systematics in the BAF data or the SNe data. However our analysis sets limits on the extent of such systematics in the DESI-DR2 analysis. In addition, it shows that parameterized deviations from the Etherington relation do not necessarily seem to be the solution for the non-constant equation of state for dark energy implied by the DESI-DR2 results.

\section{Discussion}\label{sec:discussion}

The validity of the Etherington relation can be used to also put constraints on
the redshift evolution of the absolute magnitudes of supernova, over and above
any evolution expected either from the changing distribution for the stretch
parameter, or the evolution of the properties of host galaxies that are already
accounted for in the analysis of supernovae. There are no other direct probes of
such effects and thus the Etherington relation provides a way forward.

Let us consider a simple Taylor expansion for the absolute magnitude of
supernovae, such that
\begin{align}
M(z) = M_0 + \frac{dM}{dz} z\,.
\end{align}
In this case, the inferred $\DL^{\rm inf}(z)$ will be different from the
$\DL^{\rm true}(z)$ such that,
\begin{align}
    \frac{\DL^{\rm inf}}{\DL^{\rm true}} &= 10^{0.2\frac{dM}{dz} z} \nonumber \\
    \frac{\DL^{\rm inf}}{\DA^{\rm true}} &= (1+z)^210^{0.2\frac{dM}{dz} z}
\end{align}
In this case, the ratio of the above quantity at redshifts $z$ and $z_{\rm ref}$
will be given by
\begin{align}
    {\cal R}(z, z_{\rm ref}) &= \left(\frac{1+z}{1+z_{\rm ref}}\right)^2 10^{0.2\frac{dM}{dz} z}
\end{align}
Given the large redshift lever arm for the Union3 and the Pantheon+ sample, we
fit our results for ${\cal R}$ for these two samples with this simple model. We
obtain $\frac{dM}{dz}= 0.11 \pm 0.08$, and $\frac{dM}{dz}=0.07 \pm 0.07$,
consistent with zero for the Union3 and the Pantheon+ samples, respectively
\footnote{We have only used cosmological model independent measurements in
this analysis. This can in principle be further constrained with the addition of
other BAF data such as from SDSS DR12 which constrain $\DA$ separately.}. A
similar test was performed in \cite{Marttens2025} who found a 3.2$\sigma$
evidence for a change in the absolute magnitude compared to the local value. As 
they did not rely on ratios as done in our study, the deviation they found is
likely another manifestation of the Hubble tension, rather than an evolution of
the absolute magnitude, as we test directly in this work.

Our conclusions about the consistency of the Etherington relation are also
consistent with recent independent results presented by \cite{xing_wu_2026} who
analyzed the DESI-DR2 BAF measurements but with a different methodology compared
to ours. Their results also show that any deviations are likely due to zero
point calibration differences in SNeIa absolute magnitudes or the value of
$r_{\rm d}$ for BAF. Our results rely on ratios and are thus independent of
other systematics that may be associated with other astrophysical sources.

\begin{figure}
    \centering
    \includegraphics[width=\columnwidth]{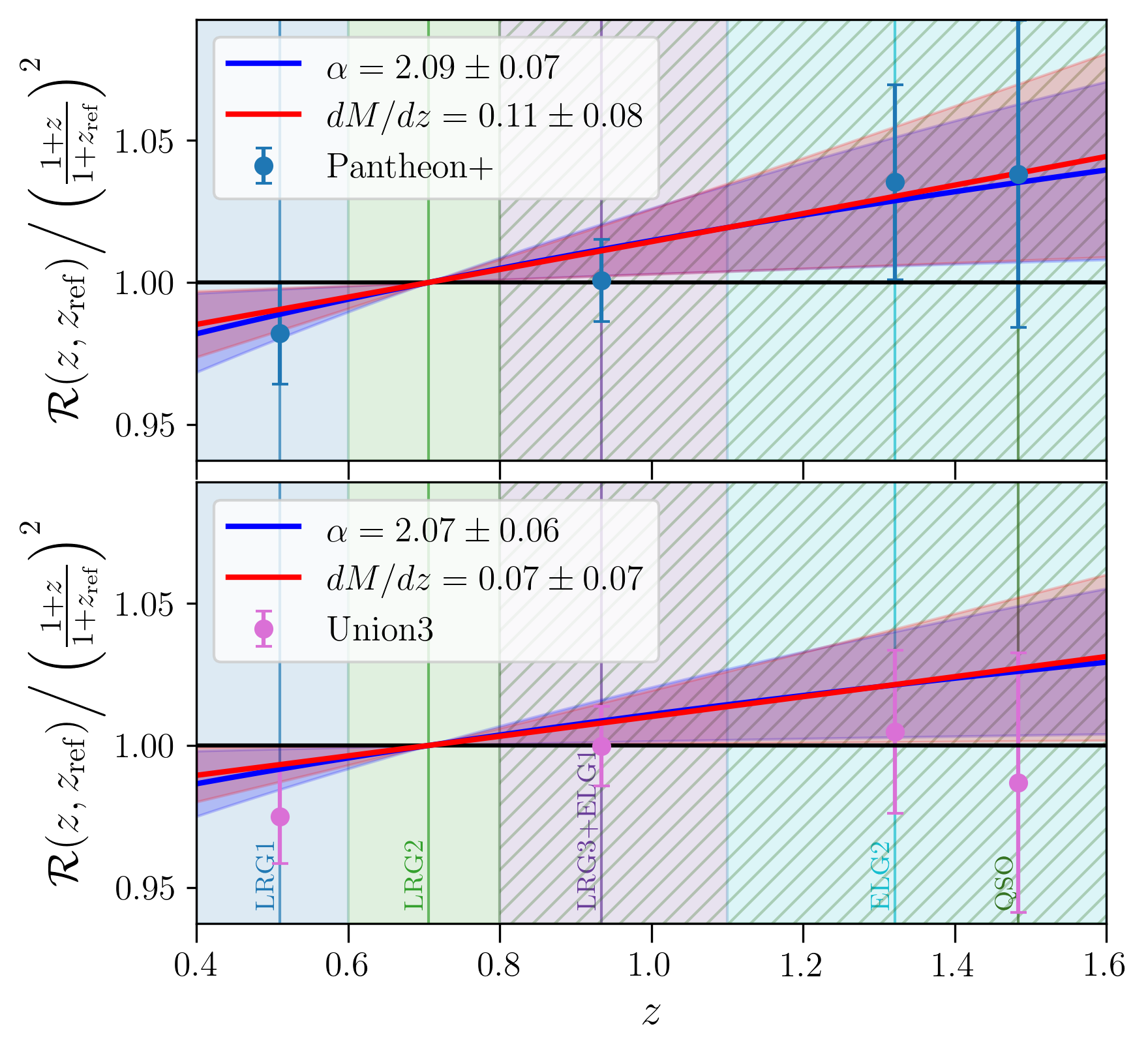}
    \caption{Deviations from the Etherington relation from the combination of DESI-DR2 with Pantheon+ (top panel) and Union3 (bottom panel) compilations, respectively. The 68 percent credible interval for this deviation with a power law model is shown as the blue shaded region, while the deviation with a model including evolution of the SNeIa absolute magnitude is shown as the red shaded region in each of the panels.}
    \label{fig:eth_evolution_panels}
\end{figure}

\begin{table}[]
\centering
\caption{Statistical consistency of the Etherington relation obtained from DESI DR2 BAF angular diameter distances and the luminosity distances from different SNeIa samples}
\label{tab:pvalues-table}
\begin{tabular}{cclll}
\cline{1-2}
\textbf{SNeIa Dataset} & \textbf{$p$-value}    &  &  &  \\ \cline{1-2}
\textbf{JLA}        & 0.11 &  &  &  \\
\textbf{Pantheon+}  & 0.59   &  &  &  \\
\textbf{Union3}     & 0.59  &  &  &  \\
\textbf{DESY5}      & 0.34 &  &  &  \\ \cline{1-2}
\end{tabular}
\end{table}

\section{Summary}

In this paper, we examined consistency between the BAF measurements from the DESI-DR2 data and the various supernova datasets that exist in the literature, namely the JLA, the Union3, the Pantheon+ and the DESY5 compilations.  Given that a combination of these datasets have shown growing significance to the possibility of dynamical dark energy, our investigation scrutinizes the consistency between data sets and hence the robustness of these conclusions. We examined these data in the context of the Etherington relation also referred in the literature as the Cosmic Transparency test, or the distance duality test. We summarize our results below.

\begin{itemize}
    \item We computed the ratio of the inferred angular diameter distance from the BAF measurements at the effective redshifts from DESI DR2 LRG1 and LRG2 samples, ELG2 sample and quasars to that at a fiducial redshift of $z=0.706$ measured from a separate combination of LRG3 and ELG1 sample. This allowed us to eliminate any dependence on the model dependent value of the sound horizon at the redshift drag epoch.
    \item We fit the distance modulus as a function of redshift using the different SNe compilation in the literature in a manner which is independent of the cosmological model. For this purpose, we fit the residual between the distance moduli implied by the data and a fiducial cosmological model with a quadratic polynomial.
    \item We computed the ratio of the luminosity distance at these same effective redshifts to that at redshift $z=0.706$, by propagating the inference from our model inferred distance moduli. The ratio of the luminosity distances were thus measured independent of the absolute magnitude of the SNe.
    \item We compared the ratio of the luminosity distances to that of the angular diameter distances to test for the $(1+z)^2$ dependence expected from the Etherington relation.
    \item For all the SNe compilations we tested, we found a remarkable statistical consistency with the Etherington relation.
    \item For the Union3 sample and the Pantheon+ sample, which span a large redshift range, the deviation from the Etherington relation of the form $(1+z)^\alpha$ was constrained such that $\alpha=0.09\pm0.07$ and $0.07\pm0.06$, respectively. Both results are statistically consistent with $\alpha=0$.
    \item If the Etherington relation is assumed to be valid, we constrained the maximum change in the unmodelled absolute magnitude of SNeIa to be $0.11\pm0.08$ and $0.07\pm0.07$, respectively for the two samples.
\end{itemize}

Given the remarkable consistency of the luminosity distance and the angular
diameter distance measurements, we see no evidence from this perspective to
question the DESI DR2 results based on the combination of these data sets. As
long as the supernovae absolute magnitudes are entirely marginalized over as
done in the DESI DR2 fiducial analyses, our results demonstrate the
appropriateness of combining the two data sets. To be absolutely immune to any
evolution of the absolute magnitude of SNeIa, a simple phenomenological model
could be applied and marginalized over as done in the discussion section, with
the prior of demanding consistency with the Etherington relation. We intend to
explore this direction further in future work.

\begin{acknowledgements}
The authors thank attendees of the various editions of the Pune-Mumbai Cosmology
and Astro-Particle (PMCAP) meetings for discussions related to this project. The
authors are also thankful to Divya Rana, Shreya Mukherjee, Khushi Lalit, Shubham
Sati for discussions during this project. SA was supported by the Department of Atomic Energy, Government of India, under Project Identification Number RTI-4012. 
\end{acknowledgements}

\section*{Data Availability}

The BAO distance measurements used in this work are from the public
DESI DR2 data release \citep{DESI_DR2}. The SNeIa compilations are
publicly available: JLA \citep{Betoule_2014}, Pantheon+ \citep{Brout_2022},
Union3 \citep{Rubin2023}, and DESY5 \citep{DESY5}.

\bibliography{refs-short}

\end{document}